\documentclass[preprint,showkeys,showpacs,preprintnumbers,amsmath,amssymb]{revtex4}

\usepackage{graphicx}
\usepackage{dcolumn}
\usepackage{bm}

\begin{document}
\newcommand{\be}{\begin{equation}}
\newcommand{\ee}{\end{equation}}
\newcommand{\ber}{\begin{eqnarray}}
\newcommand{\eer}{\end{eqnarray}}
\newcommand{\mean}[1]{\left\langle #1 \right\rangle}
\newcommand{\abs}[1]{\left| #1 \right|}
\newcommand{\set}[1]{\left\{ #1 \right\}}
\newcommand{\la}{\langle}
\newcommand{\ra}{\rangle}
\newcommand{\lb}{\left(}
\newcommand{\rb}{\right)}
\newcommand{\norm}[1]{\left\|#1\right\|}
\newcommand{\RA}{\rightarrow}
\newcommand{\eps}{\varepsilon}
\newcommand{\tNN}{\tilde{\mathbf{X}}_n^{NN}}
\newcommand{\NN}{\mathbf{X}_n^{NN}}

\title{Noise estimation by use of neighboring distances in Takens space and its
applications to stock market data}
\author{Krzysztof Urbanowicz}
 \email{urbanow@mpipks-dresden.mpg.de}
\affiliation{Max Planck
Institute for the Physics of Complex Systems\\
 N\"{o}thnitzer Str. 38\\ D--01187 Dresden, Germany}

\author{Janusz A. Ho{\l}yst}
 \email{jholyst@if.pw.edu.pl}
\affiliation{Faculty of Physics and Center of Excellence Complex
Systems Research \\Warsaw University of Technology \\
Koszykowa 75, PL--00-662 Warsaw, Poland}
\date{\today}

\begin{abstract}
We present a method that uses  distances between nearest neighbors
in Takens space to evaluate a level of noise. The method is valid
even for high noise levels.  The method has been verified by
estimation of  noise levels in several chaotic systems. We have
analyzed the noise level for Dow Jones and DAX indexes and we
have found that the noise level ranges from 25 to 80 percent of
the signal variance.
\end{abstract}


 \maketitle

\section{Introduction}
 \par  It is a common case that observed data  are contaminated
 by a noise (for a review of methods of
 nonlinear time series analysis see
 \cite{kantzschreiber,abarbanel}). The presence of noise can substantially affect invariant system parameters as a dimension,
     entropy or Lapunov exponents. In fact Schreiber
     \cite{Schreiber1} has shown that even $2\%$ of noise can make a dimension
     calculation misleading. It follows that the assessment of the noise level can
     be crucial for estimation of system invariant parameters.
        Even after performing a noise reduction one is interested to evaluate the noise
        level in the cleaned data. In the experiment the noise is often regarded
 as a measurement uncertainty which corresponds to a random variable added to
 the system temporary state or to the experiment outcome. This kind of noise is
 usually called  the {\it measurement} or the {\it additive} noise.
 Another case is the noise influencing the system dynamics, what corresponds to the Langevine
 equation and can be called the {\it dynamical} noise. The second case is more difficult to
 analyze because the
 noise acting at moment
 $t_0$
 usually changes the trajectory for $t>t_0$.
 It follows that there is no clean trajectory and instead of it an $\epsilon$-shadowed
 trajectory occurs \cite{Farmer}. For real data a signal (e.g. physical experiment data
    or economic data) is subjected to the mixture of both
    kinds of noise (measurement and dynamical).
   \par
 Schreiber has developed a method of noise level estimation \cite{Schreiber1}
  by evaluating the influence of noise on the correlation
 dimension of  investigated system. The Schreiber method is valid for
 rather small gaussian measurement noise and needs values of the
 embedding dimension $d$, the embedding delay $\tau$ and
  the characteristic dimension $r$ spanned by the system dynamics.
 \par Diks \cite{Diks} investigated properties of correlation integral with the gaussian kernel in the
  presence of noise.  The Diks method makes use of a fitting function for correlation integrals
   calculated from
 time series for different thresholds $\eps$. The
 function depends on system variables $K_2$ (correlation entropy), $D_2$ (correlation dimension), $\sigma$ (standard noise deviation)
  and a normalizing constant $\Phi$. These four variable are estimated
  using the least squares fitting. The Diks method \cite{Yu} is valid for a noise level up
 to $25\%$ of signal variance and
  for various measurement noise distributions.
  The Diks's method needs optimal values of the embedding dimension $d$,
   the embedding delay $\tau$ and the maximal threshold $\eps_c$.
 \par
 Hsu et al. \cite{Hsu} developed a method of noise \textit{reduction} and they used this method for noise level estimation. The method explored the
 local-geometric-projection principle  and is useful for various noise distributions but rather small noise
 levels. To use the method one needs to choose a number of
 neighboring points to be regarded, an appropriate number of iterations as well as  optimal parameters
 values $d$ and $\tau$.
\par
 Oltmans et al. \cite{Oltmans} considered influence of noise on
 the probability density function $f_n(\eps)$ but they could take into account only a small measurement
 noise. They used a fit of $f_n(\eps)$ to
the corresponding function which was found for small $\eps$. Their
fitting function is similar to the
 probability density distribution that we receive from correlation
 integrals
 $\frac{1}{N^2} DET_n(\eps)$. The method needs as input parameters values of $d$,
 $\tau$ and
 ${\eps}_c$.
  \par In Ref.~\cite{UrbNTS03} we presented a method of noise level estimation by
  coarse-grained correlation entropy (NECE). The crucial point of this method is fitting of a proper function to the estimated correlation
entropy.  This method does not demand any input parameters like
the embedding dimension d or the embedding delay $\tau$. The
minimal and maximal values of the threshold parameter $\epsilon$
can be automatically estimated. The NECE method  will be used
further as the reference method.
\par In this paper we present another method for evaluation of a noise level. The
method makes use of neighboring distances in the embedding space
(NEND) and will be introduced in the next section. In the further
section we show  an application of this method to stock market
data. Although it is a common believe that the stock market
behaviour is driven by stochastic processes
\cite{voit,Buchound,Mantegna} it is difficult to separate
stochastic and deterministic  components of market dynamics. In
fact the deterministic fraction  follows usually from nonlinear
effects and can possess  a non-periodic or even chaotic
characteristic \cite{Peters,Holyst}. With the help of the NEND and
NECE methods \cite{APFA4} we try to demonstrate that stock market
data are not purely stochastic and  a deterministic part can be
sometimes dominant.

\section{Method of noise estimation by  use of neighboring distances in Takens space (NEND)}
\par Let $\{x_i\}$ where $i=1,2,...,N$ be a time series and
$\mathbf{x}_i=\{x_i,x_{i+\tau},...,x_{i+(d-1)\tau}\}$ a
corresponding $d$-dimensional vector constructed in the embedded
space  where $d$ is an embedding dimension and $\tau$ is an
embedding delay. The method is based on the assumption that the
minimal distance between nearest neighbors is described by the
standard deviation of noise. The nearest neighbor is found using
the Euclidian norm i.e. the distance is measured using the
following formula
 \begin{equation} dis_{i,j}=\sqrt{\lb x_i-x_j\rb^2+\lb
x_{i-\tau}-x_{j-\tau}\rb^2+\ldots+\lb x_{i-\lb
d-1\rb\tau}-x_{j-\lb d-1\rb\tau}\rb^2}.
\end{equation} The nearest
neighbor of the vector $\mathbf{x}_n$ is the vector $\mathbf{x}_j$
such that \begin{equation} \set{\mathbf{x}_j:\forall_k
\mathbf{x}_k,(k\neq j,n),dis_{n,j}\leq dis_{n,k}}
\label{eq.NONNdef}.\end{equation}
\par We will assume that the distance between  the vector $\mathbf{x}_n$
and its  nearest neighbor
 ($dis_{n}^{NN}$) is calculated in a  large embedding dimension
$d>>1$.
\par For linear systems without a noise the minimal distance between nearest neighbors should
decrease with an increasing number of data in time series and for
$N\RA \infty$ this distance will tend to zero  since the
trajectory reaches the final periodic orbit. For deterministic
chaotic systems such minimal distances depend on the system
entropy but they also tend to zero for an infinite number of data
when the trajectory densely fills the chaotic attractor.
\par In the case when we add to an observed deterministic  trajectory a Gaussian non-correlated noise
the corresponding minimal distance  $dis_{n}^{NN}$ is increased.
For a large value of the embedding dimension $d>>1$ the distances
can be estimated as a standard deviation of a superposition of
$2d$ independent stochastic variables, i.e.
\begin{equation} dis_n^{NN}\approx \sqrt{2d}\sigma,
\label{eq.disnNN}\end{equation} where $\sigma$ is the standard
deviation of a noise added to the signal.
\par The approximation~(\ref{eq.disnNN}) is valid only in  limits  of very
long time series and a large embedding dimension $d$. If we
generate  surrogate data $\set{surr_i}$ by the random shuffling of
the original data \cite{Holyst} then this kind of surrogates
preserves mean, variance and  histogram but removes any
determinism in data. The minimal distance between nearest
neighbours calculated in an embedded space for surrogate data
should be proportional to standard deviation of data.
\par Now let us define the Noise-To-Signal ratio as the proportion of $\sigma$ to the
standard deviation of data $\sigma_{data}$ \begin{equation}
NTS=\frac{\sigma}{\sigma_{data}}.\end{equation}
\par In the first step of the method we calculate all distances between nearest
neighbors in  the original and in the surrogate data. Then we
search for the smallest distance for each data set:
$dis_{min}^{NN}=\min_n\set{dis_n^{NN}}$ and
$dis_{min}^{surr,NN}=\min_n\set{dis_n^{surr,NN}}$. Using the
approximation~(\ref{eq.disnNN}), i.e. the linear dependence of the
distance $dis_{min}^{NN}$ on the noise level, we can introduce the
output parameter of the method $ADET_d$ \cite{urbanti04}, which is
related to the Noise-To-Signal ratio ($NTS$) as follows:
\begin{equation} NTS\approx ADET_d\equiv
\frac{dis_{min}^{NN}}{\mean{dis_{min}^{surr,NN}}}
\label{NTSdisnNN}.\end{equation} Here we denoted
$\mean{dis_{min}^{surr,NN}}$ as an average of $m$ realizations of
the surrogate data ($m$ appears as the parameter of the method).
\section{Noise estimation: examples and application to stock market data}
\par The NEND method described in the previous section is  very simple to use.
A drawback of this method  is a large error of the estimated noise
level for short time series and too low embedding dimensions $d$.
The estimation error increases if we take a smaller number of
random surrogates data that are used for the averaging
formula~(\ref{NTSdisnNN}). In the Table~\ref{tab:tab.1} examples
of noise estimation by the  NEND method are presented in
comparison to results of NECE method (see \cite{UrbNTS03}). The
estimation error of the NEND method is based on a standard
deviation of temporary values of $ADET_d$ for different
realizations of  surrogate data. One can see that the NEND method,
despite its simplicity, works quite well for considered cases.
Although the  NECE method gives better accuracy as compared  to
the NEND method but the first method is  much more sophisticated
and difficult for computer  implementation. CPU times needed by
computers are comparable for both methods.
\begin{table}
\caption{\label{tab:tab.1} Results of the noise level estimation
for different systems. In the case of NEND method we used $d=9$,
$m=20$ and $N=3000$.}
\begin{ruledtabular}
\begin{tabular}{ccccc}
System&$NTS$&$\sigma$&estimated $\sigma$ using NEND &estimated $\sigma$ using NECE\\
\hline

  Henon & $0$& 0 & $0.0\pm 0.05$& $-0.0023\pm 0.0001$ \\
  Henon &$0.09$& 0.1 & $0.05\pm 0.06$& $0.1\pm 0.0007$ \\
  Ikeda &$0.1$& 0.07& $0.05\pm 0.04$& $0.07\pm 0.0005$ \\
  Lorenz &$0.43$& 4 & $3.7 \pm 0.5 $& $4.4\pm 0.4$\\
  Lorenz &$0.85$& 15.7 & $16.6 \pm 1$& $14.9\pm 0.08$ \\
  Lorenz &$0.96$& 30 & $32.8\pm 1.5$& $30.5\pm 0.7$\\
  Roessler &$0.33$& 7.4&$4.7\pm 1.2$ & $6.4\pm 0.8$\\
  Roessler &$0.84$& 15 & $14.46\pm 0.8$ & $14.7\pm 1.1$\\
  Roessler &$0.93$& 33.4& $37\pm 1.8$& $33.8\pm 1$\\
\end{tabular}
\end{ruledtabular}
\end{table}
\par
\begin{figure}
\begin{center}
\includegraphics[scale=0.5, angle=-90]{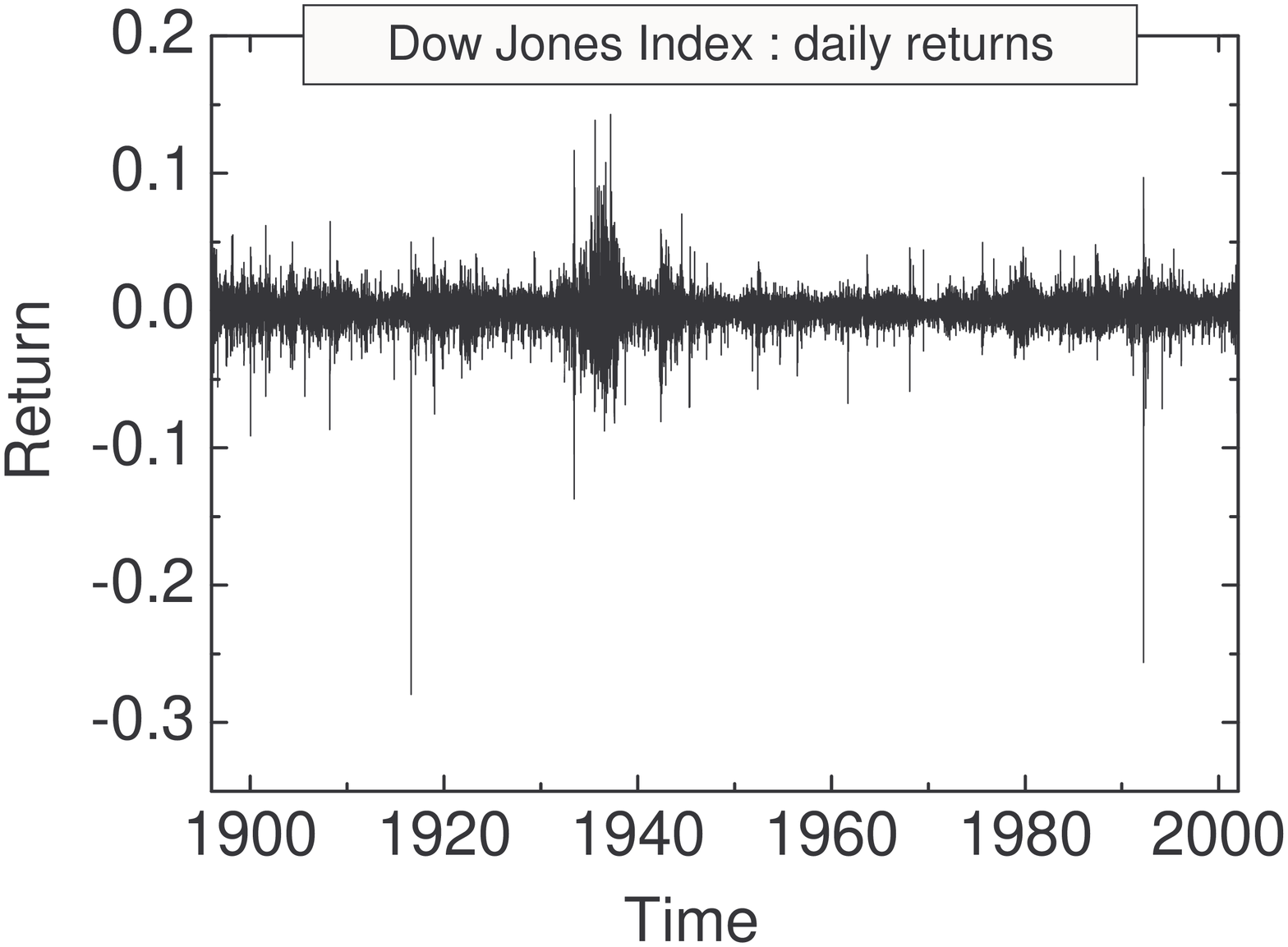}
\end{center}
\caption{\label{fig.tsDJIA} Plot of daily returns of Dow Jones
Index (1896-2002).}
\end{figure}

\begin{figure}
\begin{center}
\includegraphics[scale=0.5, angle=-90]{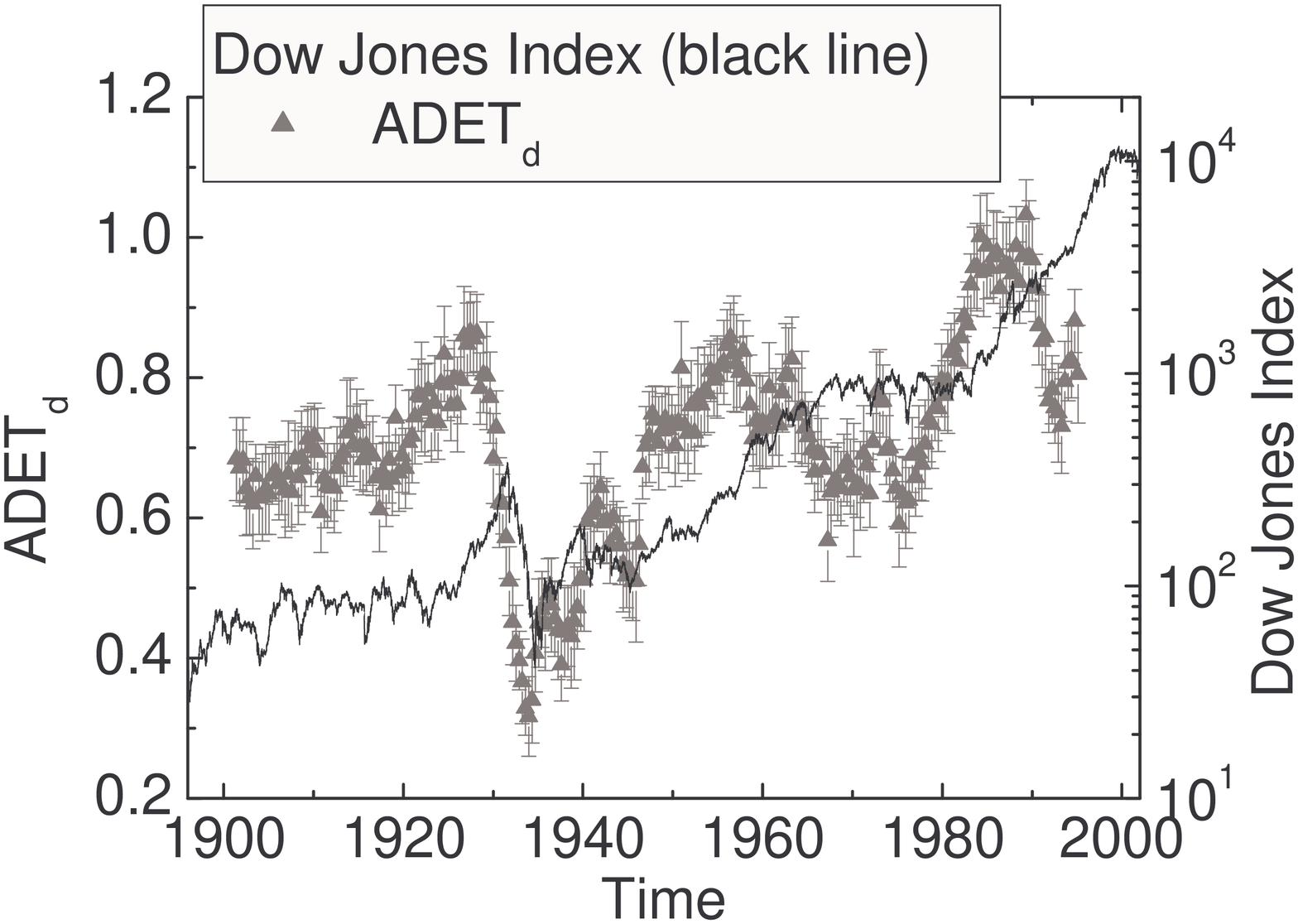}
\end{center}
\caption{\label{fig.DJIAADET} Plot of noise levels $ADET_d$ for
$d=9$ calculated with the NEND method and the value of Dow Jones
Index (1896-2002).}
\end{figure}

\begin{figure}
\begin{center}
\includegraphics[scale=0.5, angle=-90]{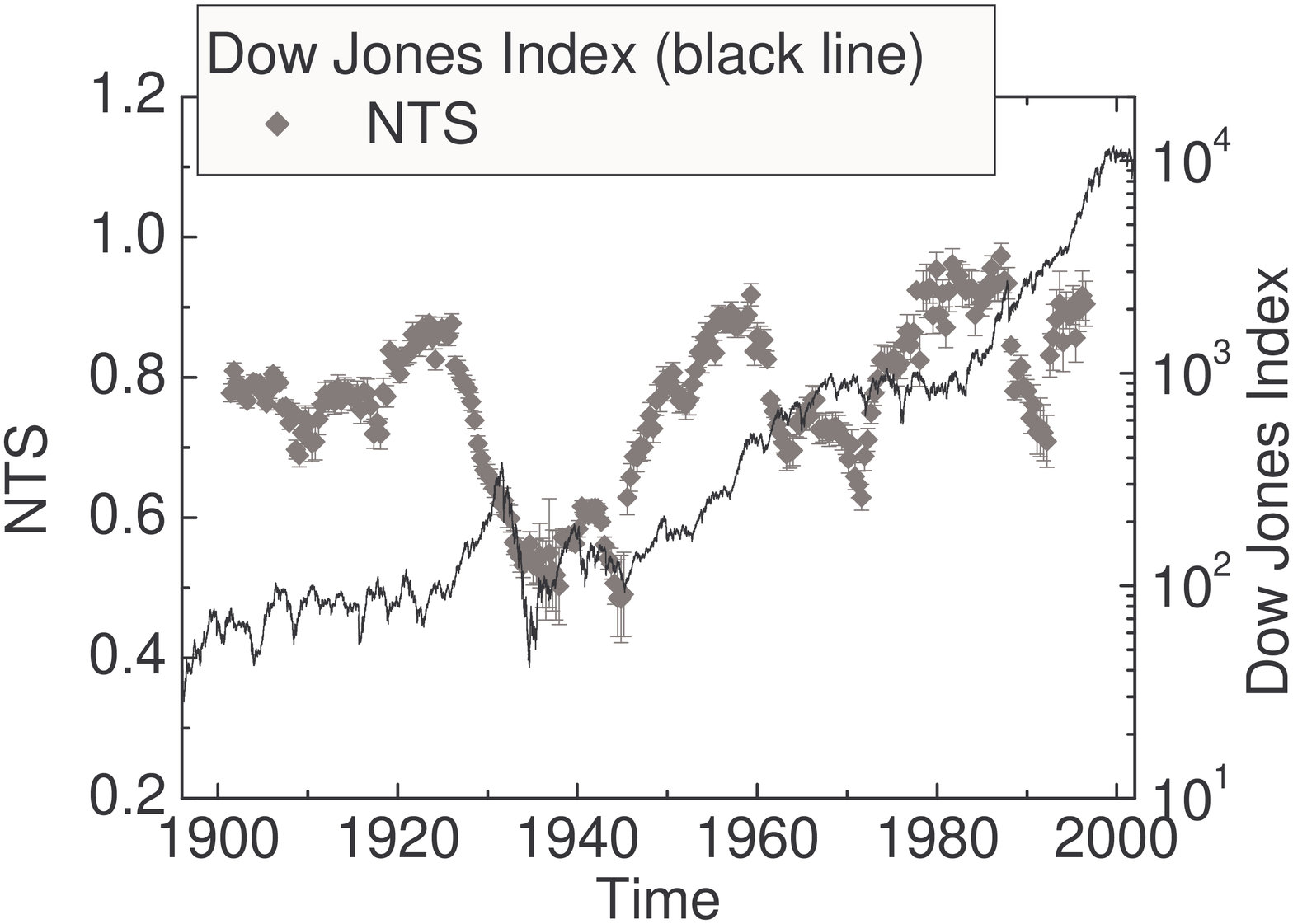}
\end{center}
\caption{\label{fig.DJIANTS} Plot of noise levels $NTS$ calculated
with the NECE method and the value of Dow Jones Index
(1896-2002).}
\end{figure}

\begin{figure}
\begin{center}
\includegraphics[scale=0.5, angle=-90]{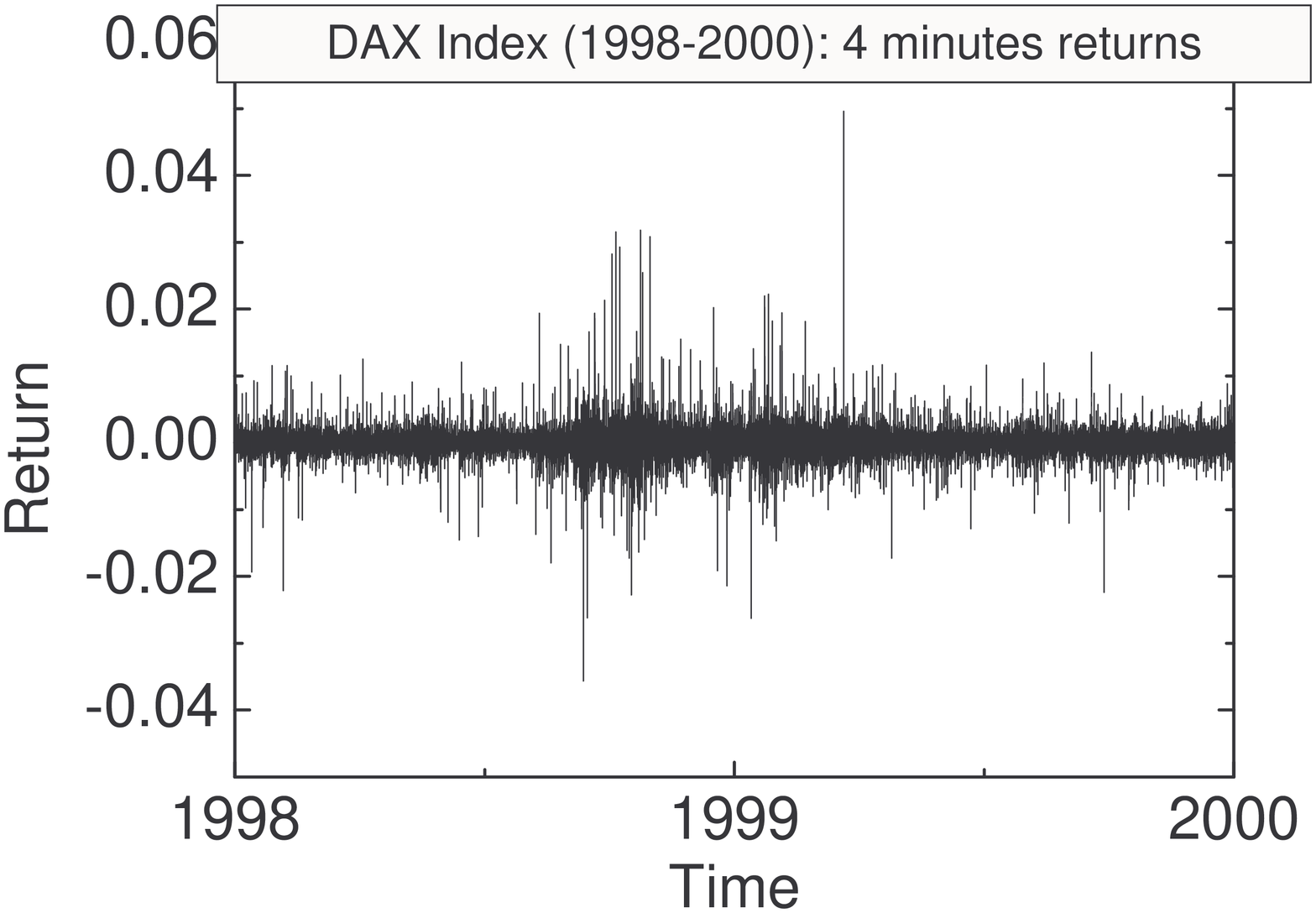}
\end{center}
\caption{\label{fig.tsDAX} Plot of 4 minutes returns of DAX Index
(1998-2000).}
\end{figure}

\begin{figure}
\begin{center}
\includegraphics[scale=0.5, angle=-90]{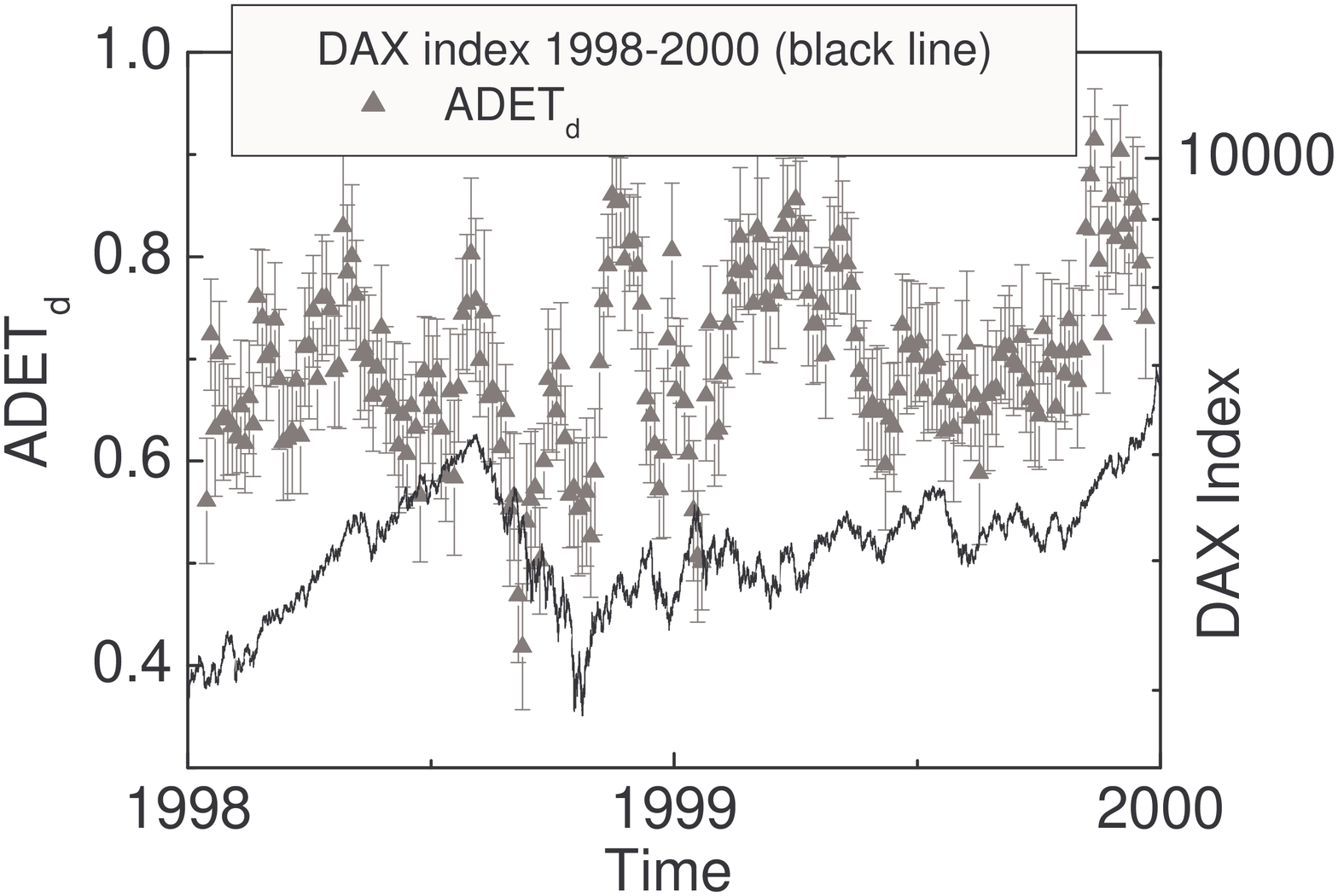}
\end{center}
\caption{\label{fig.DAXADET} Plot of noise levels $ADET_d$ for
$d=9$ calculated with the NEND method and the value of DAX Index
(1998-2000).}
\end{figure}

\begin{figure}
\begin{center}
\includegraphics[scale=0.5, angle=-90]{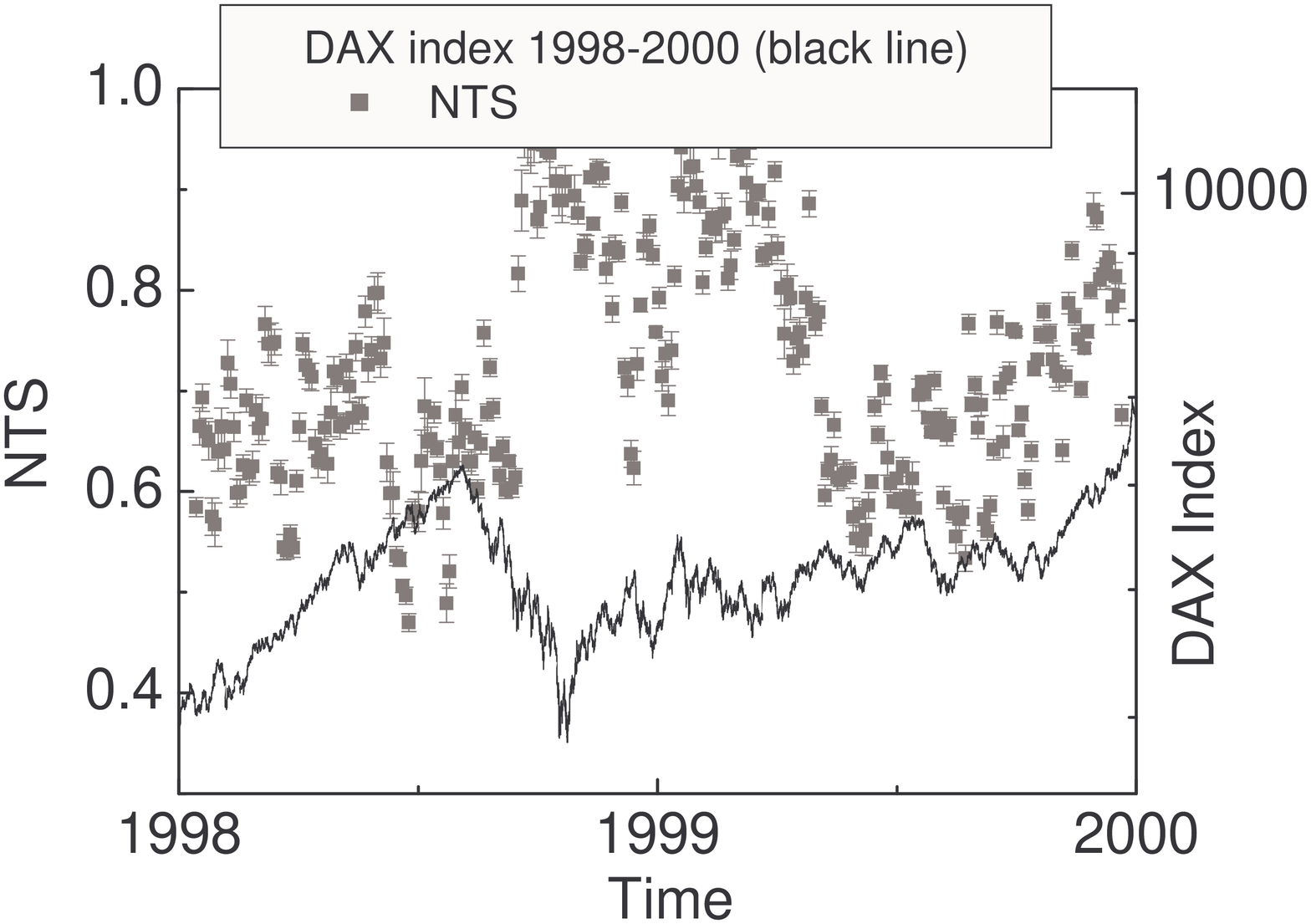}
\end{center}
\caption{\label{fig.DAXNTS} Plot of noise levels $NTS$ calculated
with the NECE method and the value of DAX Index (1998-2000).}
\end{figure}

 The both methods   were applied to evaluate the noise levels
in stock market data. Here we present results for Dow Jones
Industrial Average (DJIA) during the time period 1896-2002 (daily
returns, see Fig.~\ref{fig.tsDJIA}) and DAX (German Stock Market
Index) during the  time period 1998-2000 (4 minutes returns, see
Fig.~\ref{fig.tsDAX}). Returns $x_n$ are defined as
\begin{equation} x_n=\ln\lb \frac{P_n}{P_{n-1}}\rb,\end{equation}
where $P_n$ is the value of an index at the time $n$. Noise levels
for both indexes are in the range $NTS\approx 0.5-0.9$ as one can
see in
Figs~(\ref{fig.DJIAADET},\ref{fig.DJIANTS},\ref{fig.DAXADET},\ref{fig.DAXNTS}).
It follows  that considered  stock market data are not purely
stochastic because the percent of determinism ranges
$(1-NTS^2)\cdot 100\%\approx 20-75\%$ and the stochastic part is
about $25-80\%$. In Figs~(\ref{fig.DJIAADET},\ref{fig.DAXADET}) we
present noise levels $ADET_9$ calculated with the NEND method for
DJIA and DAX indexes respectively. For the comparison in
Figs~(\ref{fig.DJIANTS},\ref{fig.DAXNTS}) we show noise levels
$NTS$ calculated with the NECE method. The NEND method and the
reference method NECE give similar results. Since  both methods
are different approaches thus similar results suggest good
accuracy of both methods. Some differences in noise levels
estimations between both methods appear in the periods of
increased volatility (1930-1940 for DJIA and from August to
November of 1998 for DAX), what suggests that extreme events have
different impact in both methods. We think that the NECE method
gives more relevant results in high volatility regions and the
NEND method underestimates the noise level in such cases.
\section{Conclusions}
\par In conclusion we have developed a new method of noise level estimation from time series.
The method makes use of the minimal distance between nearest
neighbors in Takens space. The method has been tested for several
systems and it has  brought similar results to the method
described in  Ref.~\cite{UrbNTS03} but it is much easier for
computer implementation. The application of the method to stock
market data gives the percent of noise ranging from 25 to 80 $\%$
of signal variance.


\begin{thebibliography}{01}
\bibitem{kantzschreiber} H. Kantz and  T. Schreiber, \textit{Nonlinear Time Series Analysis} (Cambridge University Press, Cambridge, 1997).
\bibitem{abarbanel} H.D.I. Abarbanel, \textit{Analysis of Observed Chaotic Data} (Springer, New York, 1996).
\bibitem{Schreiber1} T. Schreiber, Phys. Rev. E \textbf{48(1)},1s3(4) (1993).
\bibitem{Farmer} J. D. Farmer and J.J. Sidorowich, Physica D \textbf{47}, 373-392 (1991).
\bibitem{Diks} C. Diks, Phys. Rev. E \textbf{53(5)},4263(4) (1996).
\bibitem{Yu} Dejin Yu, M. Small, R.G. Harrison and C. Diks, Phys. Rev. E \textbf{61(4)},3750(7) (2000).
\bibitem{Hsu} R. Cawley and Guan-Hsong Hsu, Phys. Rev. A \textbf{46(6)}, 3057 (1992).
\bibitem{Oltmans}H. Oltmans and P. J.T.Verheijen, Phys. Rev. E \textbf{56(1)},1160(11) (1997).
\bibitem{UrbNTS03} K. Urbanowicz and J. A. Ho{\l}yst,
Phys. Rev. E \textbf{67}, 046218 (2003).
\bibitem{voit} J.Voit, \textit{The
Statistical Mechanics of Financial Markets}, (Springer-Verlag
2001).
\bibitem{Buchound} J.P. Bouchaud, M. Potters, \textit{Theory of financial risks - from
statistical physics to risk management}, (Cambridge University
Press 2000).
\bibitem{Mantegna} R.N. Mantegna, H.E. Stanley, \textit{An
Introduction to Econophysics. Correlations and Complexity in
Finance}, (Cambridge University Press 2000).
\bibitem{Peters} E.E. Peters, \textit{Chaos and Order in the
Capital Markets. A new view of cycle, Price, and Market
Volatility}, (John Wiley $\&$ Sons 1997).
\bibitem{Holyst} J.A. Ho{\l}yst, M. \.Zebrowska and K. Urbanowicz, European Physical Journal B \textbf{20}, 531-535 (2001).
\bibitem{APFA4} K. Urbanowicz,  and J. A. Ho{\l}yst, Physica A
\textbf{344}, 284-288 (2004).
\bibitem{urbanti04} K. Urbanowicz, H. Kantz and J. A. Ho{\l}yst,
"Anti-deterministic behavior of discrete systems that are less
predictable than noise", in press in Physica A,
arXiv:cond-mat/0408429 (2004).
\end{thebibliography}
\end{document}